# A Conceptual Model for Holistic Classification of Insider

Ikuesan R. Adeyemi*,    Shukor Abd Razak    Mazleena Salleh
raikuesan2@live.utm.my    shukorar@fsksm.utm.my    mazleena@fsksm.utm.my
*Information Assurance and Security research group,
Faculty of Computing, Universiti Teknologi, Malaysia*

**Abstract**

The process through which an insider to an organization can be described or classified is lined within the orthodox paradigm of classification in which an organization considers only subject with requisite employee criterion as insider to that organization. This is further clouded with the relative rigidity in operational security policies being implemented in organizations. Establishing investigation process in instances of misuse occurrence and or ascertaining efficiency of staff member using such archaic paradigm is maligned with endless possibilities of uncertainties. This study therefore proposes a holistic model for which insider classification can be crystallized using the combination of qualitative research process, and analysis of moment structure evaluation process. A full comprehension of this proposition could serve as a hinge through which insider misuse investigation can be thoroughly carried out. In addition, integrating this paradigm into existing operational security policies could serve as a metric upon which an organization can understand insider dynamics, in order to prevent misuses, and enhance staff management.

## 1. INTRODUCTION

The 2010 Cyber Security watch survey (Keeney & Rogers, 2010) with over 500 respondents uncovers a hidden reality that shows that considerable percentage of cybercrimes are committed by neither an insider nor an outsider, thus tagged unknown. Such attacks includes unauthorized access to/use of information, system, and network; intentional exposure of private/sensitive information. Whilst some attacks could be deniably non-insider origin such as spyware, others are arguably attacks that emanates from within the organization. This level of relatively high unknowns could be attributed to the relative ambiguity and anonymity inherent in the traditional information security policies, which views an insider from the paradigm of subject with legitimate access right, and clearance privileges only. This paradigm of defining insider based on subject-object relationship in isolation does not reveal the reality of operational process of human behaviour. Furthermore, it neglects the sociological paradigm of human interaction (psychosocial attributes), while it exposes the inherent vulnerabilities in systems (G. B. Magklaras & Furnell, 2002). The implication of these ambiguity and obscurity range from the lack of detection and inaccurate identification of insider misuse, to anonymity inclusion; which hinders the possibilities of investigation, while either tarnishing reputation of organization or inhibiting operational efficiency amongst others. This paper therefore introduces a conceptual model for insider description, which can be adapted into existing policies, to project clarity in meaning, and clarification of an insider. This is however not similar to the usual practice of defining insider based on intent, in which an insider is classified according to benign intention, malicious intention or erroneous action. To the best of our knowledge, this is the first time such paradigm is empirically modelled to classify insider, from the longitudinal and vertical axis which presupposes that an insider to an organization scopes beyond traditional paradigm, especially where insider misuse investigation is called into play. This paper thus explores the dimensions of users, classified as insider to an organization. It begins with a brief overview of existing research on insider taxonomy, followed by a theoretical bases and a description of the empirical procedures adopted for this study. The result of the study is then presented followed by a detailed discussion of the implication of the result.

## 2. RELATED WORKS

Insider taxonomy of geographical delineation (Neumann, 2010) identifies subjects based on logical and physical presence to an organization infrastructure as graphically represented in Figure 1. As shown in the graphical representation, Neumann, (2010) identified four descriptive classifications: classes A, B, C, & D; which identifies an insider as someone with physical and or logical privilege to use a particular system/facility. Other literatures (Magklaras & Furnell, 2004; Magklaras, Furnell, & Papadaki, 2011; Neumann, 1999; Sarkar, 2010; Salem, Hershkop, & Stolfo, 2008) adopted the description of an insider as a subject with legitimate access right to an object.

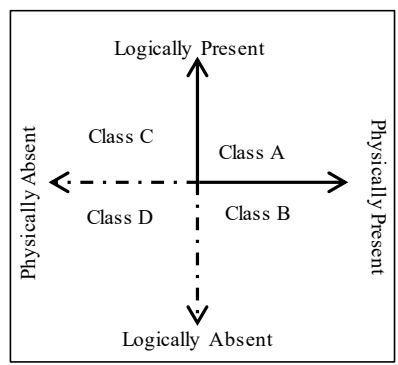

**Figure 1**: Geographical taxonomy of insider

These literatures are targeted at delineating benign insider from a malicious insider with various other taxonomies that examines the description of a malicious insider. Thus the literatures neglects the primary criteria for classification – insider. A thorough anatomy of who an insider really is spans beyond the ideal paradigm presented in these literatures. However, Adeyemi, et al., (2013) presents the state of the art surveys on insider taxonomy. The paper identified four distinct abstractions within which an insider can be defined. This includes:

- a current employee by an organization (CE)
- a laid-off employee (LE)
- a contract employee (CS)

- an affiliates to a current employee/contract staff (NCE)

Though the paper attempted to generically classify insider without prejudice to malicious taxonomy, it however failed to present an empirical validation to substantiate the abstractions. This study builds on the abstraction in Adeyemi et al., (2013), by extracting the psychosocial attributes using output matrix, series of self-administered questionnaire, and a structural modelling evaluation process. The proceeding section details the methodology and process implored in this study.

## 3. THEORETICAL DESCRIPTION OF PROPOSED INSIDER TAXONOMY

The four abstractions in (Adeyemi et al. 2013) is plotted on an auto-covariance matrix/description to derive an output matrix depicted in Table 1. This output matrix depicts inter-relational inference which results in a forty-dimension description that form various classes of personnel, ranging from current employee, contract staffs, collaborators and other stakeholders. These classes reveal the psychosocial tendencies in which human interaction is described, which considers it erroneous, to rule out the influence, affluence, and effect each classes could generate.

Table 1: Auto-covariance (outcome) Matrix of Insider

| Human Interaction paradigm | Subjects-Object Paradigm | | | |
|---|---|---|---|---|
| | CE | LE | CS | NCE |
| CE | $A_1B_1$ | $A_2B_1$ | $A_3B_1$ | $A_4B_1$ |
| LE | $A_1B_2$ | $A_2B_2$ | $A_3B_2$ | $A_4B_2$ |
| CS | $A_1B_3$ | $A_2B_3$ | $A_3B_3$ | $A_4B_3$ |
| NCE | $A_1B_4$ | $A_2B_4$ | $A_3B_4$ | $A_4B_4$ |
| CE-LE | $A_1B_5$ | $A_2B_5$ | $A_3B_5$ | $A_4B_5$ |
| CE-CS | $A_1B_6$ | $A_2B_6$ | $A_3B_6$ | $A_4B_6$ |
| CE-NCE | $A_1B_7$ | $A_2B_7$ | $A_3B_7$ | $A_4B_7$ |
| LE-CS | $A_1B_8$ | $A_2B_8$ | $A_3B_8$ | $A_4B_8$ |
| LE-NCE | $A_1B_9$ | $A_2B_9$ | $A_3B_9$ | $A_4B_9$ |
| CS-NCE | $A_1B_{10}$ | $A_2B_{10}$ | $A_3B_{10}$ | $A_4B_{10}*$ |

*The letters "A" and "B" are arbitrary representation with no connotative meaning to the outcome matrix

The abstractions CE, LE, CS, and NCE refers to subject in an organization, while the abstraction (also referred to as superset) CE.LE, CE.CS, CE.NCE, CS.LE, CS.NCE, and LE.NCE refers to interaction between each identified subjects. Subject-object paradigm is based on access privileges permitted to a subject on an object. Example of such is the relationship defined by the BellLa-Padular confidentiality model (Tilborg and Jajodia 2011), in which a subject is bounded by the clearance criterion to an object. Moreover, human interaction paradigm entails the possible communication between subjects of same or differing classes, over an object of group of objects. This process may involve the use of experience garnered from previous observation, deliberate communication with other subject, or even collaboration with other subjects. Example of such instance could be an IT expert who doubles as a bank cashier, in addition to being a staff to technical consultancy firm which deals with banks. Another example of such a superset could be a university staff that doubles as member of two faculties within the university, or instances where married couples are stationed in different department within same organization. Against this backdrop, it could be assert that an insider transcends the boundary of subject-object paradigm. Table 2 gives a classified description which this study sought to empirically evaluate to explicate the dimension of human interaction paradigm in insider definition.

Table 2: Classification of Insider

| Single Variant Subject | Unique Single Varian Subject | Double variant subject | Unique Double Variant Subject | Triple Variant Subject |
|---|---|---|---|---|
| CE | CE→CE | CE↔LE | CE↔CE-LE | CE↔LE-CS |
| LE | LE→LE | CE↔CS | CE↔CE-CS | CE↔LE-NCE |
| CS | CS→CS | CE↔NCE | CE↔CE-NCE | CE↔CS-NCE |
| NCE | NCE→NCE | LE↔CS | LE↔CE-LE | |
| | | LE↔NCE | LE↔LE-CS | |
| | | CS↔NCE | LE↔LE-NCE | |
| | | | CS↔CE-CS | |
| | | | CS↔LE-CS | |
| | | | CS↔CS-NCE | |
| | | | NCE↔CE-NCE | |
| | | | NCE↔LE-NCE | |
| | | | NCE↔CS-NCE | |

→ refers to a possible communication process or acquired knowledge by a subject
↔ refers to a mutual communication process

These five classifications expanded in Table 2 can be further represented using mathematically notations as shown in Equation (1) through (4).

### 3.1 Single Variant Subject Class (SVSC)

This is a subject-object paradigm. Subject in this class possess only the requisite access right and access knowledge for specific assigned responsibility in an organization.

### 3.2 Unique Single Variant Subject Class (USVSC)

Subject in this class possess knowledge formed through the interaction between two subjects in an organization, each belonging to different departments, but same access right and requisite access knowledge within the organization. Equation (1) gives a mathematical composition of SVSC extrapolated for the outcome matrix in Table 1.

$$\text{USVSC}_{(A, B)} \text{ Class} = \sum_{n=1}^{4} A_n B_n \tag{1}$$

This class forms the diagonal of the 4x4-subset matrix, of the auto-covariance matrix defined in Table 1. This class of subject shares the same probability of existence with SVSC.

### 3.3 Double Variant Subject Class (DVSC)

These refer to subjects that possess knowledge formed through the interaction between member of different access knowledge, but not necessarily different access right, vice versa. Such knowledge tends to transcend the requisite access knowledge of a singular subject. Equation (2) illustrates the mathematical representation of this class.

$$\text{DVSC}_{(A, B)} \text{ Class} = \sum_{n=2}^{4} A_1B_n - \sum_{n=3}^{4} A_2B_n - A_3B_4 \qquad (2)$$

### 3.4 Unique Double Variant Subject Class (UDVSC)

These subjects possess higher knowledge of the organization, because of multi-interaction among subjects of different departments within an organization such that at least, two of the subjects have same access right but not necessarily within same access knowledge.

$$\text{UDVSC}_{(A, B)} \text{ Class} = \sum_{n=5}^{7} A_1B_n + [\sum_{n=8}^{9} A_2B_n + A_2B_5] + \sum_{n=3}^{5} A_3B_{2n} + [\sum_{n=9}^{10} A_4B_n + A_4B_7] \qquad (3)$$

### 3.5 Triple Variant Subject Class (TVSC)

These are subjects capacitated with the knowledge accumulated from at least three subjects within an organization, each with a distinct access knowledge and or access right. Equation (4) gives the mathematical representation of this class of insider.

$$\text{TVS}_{(A, B)} \text{ Class} = \sum_{n=8}^{10} A_1B_n \qquad (4)$$

Equations (1) through (4), can be further grouped using common factor analysis. SVSC and USVSC can be called a single class, DVSC and UDVSC can be referred to as a double class, while TVSC can be called a triple class. Thus equation (1) and (2) represents a uni-variant subject class, equation (3) and (4) represents a di-variant subject class, and equation (5) represents a group of tri-variant subject class. The interaction between subjects in each group forms the cardinal of the theoretical model as shown in Figure 2.

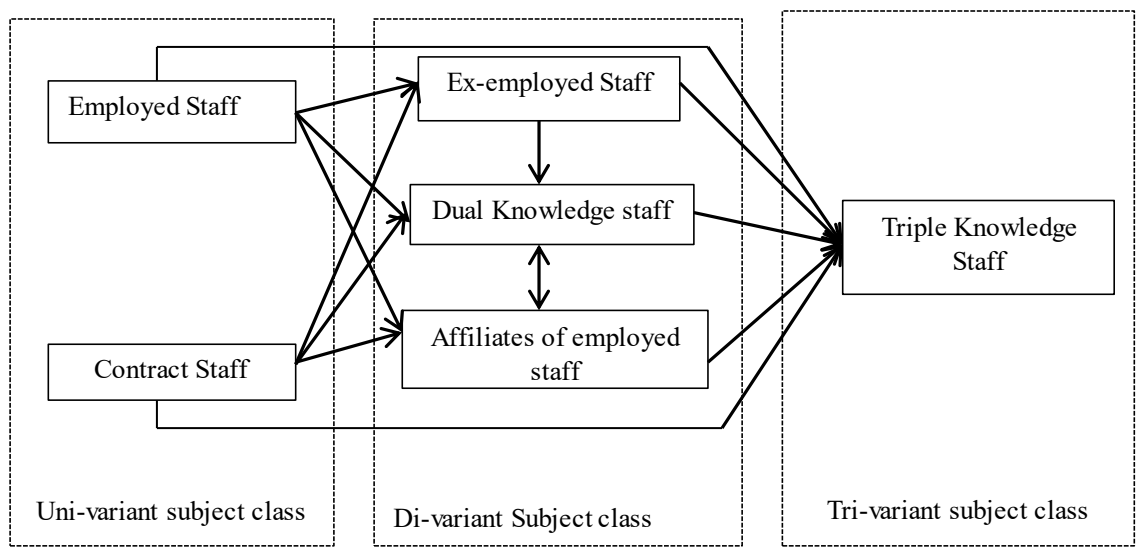

Figure 2: Theoretical Model for Insider Taxonomy

Leveraging the outcome matrix, the theoretical model shown in Figure 2 stations this study as an interactive integrated composition of the various classes of subjects considered to form the cardinal in which an insider is formed. This research thus set the paradigm of insider formation as an extension to the traditional definition of insider.

## 4. Empirical Procedure Of The Proposed Insider Description Model

This section details the quantitative factorization of the conceptual model. Self-administered questionnaire instrument was developed for this process. This study adopts the process defined in (Bartlett, Kotrlik, and Higgins 2001; Israel 1992) as shown in Equation (5)

$$n = \frac{\{(p(1-P))/\left(\frac{A^2}{Z^2}\right) + \frac{P(1-P)}{N}}{R} \qquad (5)$$

Where $n$ = sample size: $N$ = estimate of population: $P$ = estimated variance in population
$A$ = desired precision: $Z$ = confidence level: $R$ = estimated response rate.

Table 3 elaborates on the adopted selected values for each parameter, which constitutes the sample size.

**Table 3**: Sample size calculation

| Organization classification | Population variance | Precision | Confidence level | Estimate of population | Estimated response rate | Sample size |
|---|---|---|---|---|---|---|
| Public | 50% | 3% | 95% | 200 | 70% | 168 |
| Private | 50% | 3% | 95% | 100 | 60% | 81 |
| Government | 50% | 3% | 95% | 100 | 80% | 81 |

As shown in Table 3, the respondents covers three tiers of organization -private, governement, and public. The private institution represents section in observation which are owned and controlled by private individuals, such as insurance, banks and so on. The government institution represents section in investigation caseload that is controlled by the government of a nation such as the government operational centres and government administrative centres. The public institution on the other hand represents sections in investigation, which are controlled by the Government through proxies, such as the public libraries and public universities. Two public universities were selected to represent the public institution, a government operational centre is selected to represent the government institution, and a Commercial bank was selected to represents private institutions, all in two different states in Malaysia. A total of 153, 89, and 82 respondents were collected for public, government, and private institutions respectively.

The questionnaire design is structured into five-phase 34-item, self-rated semantic differential pattern to reflect the description of insider identified in existing literatures (Al-Morjan, 2010; Anderson, 1980; Hunker & Probst, 2011; G Magklaras & Furnell, 2004; Neumann, 1999; Roy Sarkar, 2010). These phases include description of employed staff, contract staff, sacked staff, affiliates, and double knowledge staffs. The pattern matrix from the analysis of the result further reveals the delineation of phases herein defined. The distribution and collection of the questionnaire spanned 7 weeks (3rd February 2013 to 25$^{th}$ March, 2013). The respondents comprise top management, middle management, administrative staff, contract staff, and technical staffs. We do not claim that this sample covers the entirety of the population of Malaysia, but serves as representative of the Malaysia workforce.

However, a 10 days pilot survey comprising 15 respondents was conducted to initially ascertain the skewness of the questionnaire. As expected the outcome of the analysis is positively skewed. Various tests are conducted on the data, such as:

- Kaiser-Meyer-Olkin (KMO): KMO is a sampling adequacy test, while Bartlett's measure is a Sphericity test. KMO varies from 0 to 1. A value of 0 indicates low correlation among variable, while a value close to 1 indicates the high correlation. High correlation indicates that the factor analysis is reliable. Recommended range of 0.5-0.7, 0.7-0.8, 0.8-0.9, and above 0.9 to be mediocre, good, great value, and superb value respectively.
- Bartlett's Measure: Bartlett's measure is a test of null hypothesis, which states that the correlation matrix is an identity matrix. This illustrates that the variables are not related, thus unsuitable for factor analysis. Values of range < 0.05 indicates that the data is suitable for factor analysis. Otherwise, the data is not suitable for structure detection.
- Missing Completely at Random (MCAR) Test: MCAR test is also referred to as Little's test. It is a test for biases in dataset with missing variables. A statistically significant result of MCAR indicates that the missing data in a dataset is biased, while an insignificant statistical result indicates a completely missing at random

## 5. Result of The Empirical Process

The percentage of male and female respondents are 38.3% and 60.5% respectively, which is synonymous with the findings in (Ashari 2012) where professional and management occupational level of male to female is 39.4% to 60.6%. A concise description of the demography of the respondents is shown in Table 4. 35.39% of the overall respondents are top management staffs.

Table 4: Synopsis of Demography of Respondents

| Gender | Sample Size % | Marital Status% | | Educational level% | | Job Experience% | |
|---|---|---|---|---|---|---|---|
| | | Married | Single | High | Low | < 4 years | >4years |
| Female | 60.5 | 70.07 | 25.85 | 74.15 | 25.85 | 36.05 | 63.95 |
| Male | 38.3 | 70.97 | 24.73 | 83.87 | 15.05 | 35.48 | 64.52 |

The rest of the section describes the process and the recommendation used in this study. Empirical measurement test such as composite reliability and validity, MCAR, are carried out on the dataset.

## 5.1 Little's Test of MCAR

This test was carried-out using the process described in Figure 3. As shown in Figure 3, expectation maximization (EM) algorithm is used. The results, as shown in Table 4, which indicates percentage of the missing values, shows that the percentage of the highest missing-value (1.5%) is less than the negligible threshold of 2%.

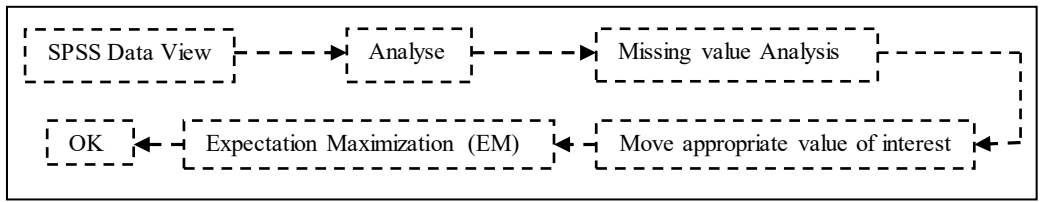

**Figure 3**: Steps to Little's MCAR Test

**Table 5**: Little's MCAR Test

| Parameters | Values |
|---|---|
| Chi-Square | 374.453 |
| Degree of Freedom (DF) | 398 |
| Sig | 0.796 |

As shown in Table 5, the result of the EM test shows it is not statistically significant (0.796), thus, indicating a randomized MCAR. Hence, the Little's MCAR test implies that data imputations can be carried-out on the dataset, to complete the missing values.

## 5.2 Data Screening Test

In order to identify the possible number of factors within which these data can be classed, we adopted the exploratory factor analysis on SPSS statistical tool. Exploratory Factor Analysis (EFA) is carried out on the data in accordance with the procedure described in Figure 4.

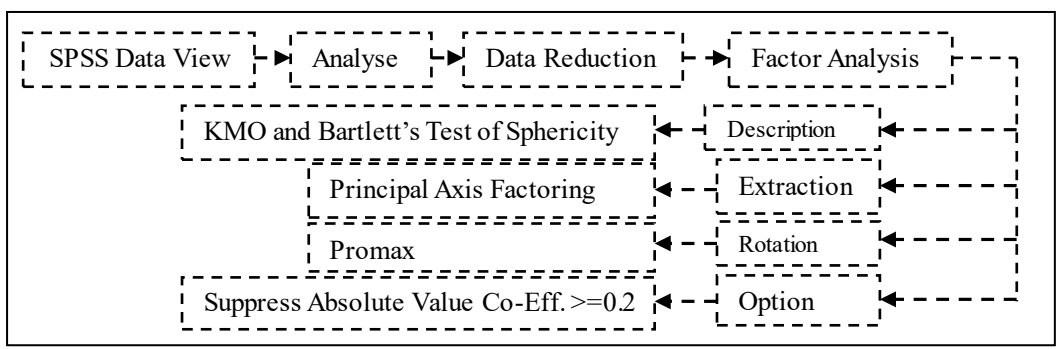

**Figure 4**: Exploratory Factor Analysis procedure in SPSS

As shown in Figure 4, KMO and Bartlett's test are carried-out on the dataset. It is used to evaluate the adequacy and appropriateness of the set of observed variable (dataset) for EFA. Table 6 shows the result of the test.

**Table 6**: KMO and Bartlett's Test

| KMO Measure of Sampling Adequacy | Bartlett's Test of Sphericity | |
|---|---|---|
| 0.913 | Approx. Chi-Square | 6.063E3 |
| | df | 561 |
| | Sig. | 0.000 |

The result presented in Table 6, shows that the dataset is adequate and appropriate for EFA. Both KMO value (0.916) and Bartlett's measure value (<0.001), indicates the appropriateness of factor analysis otherwise known as structural detection on the dataset.

### 5.3 Factor Analysis

Factor analysis is done to determine the number of underlying factors in a surveyed data, as well as to eliminate outlier variables. In order to derive statistically significant structure from the dataset, EM imputation is first carried out on the dataset to fill-up all missing data. Figure 5 shows the flow chart of the factor analysis process.

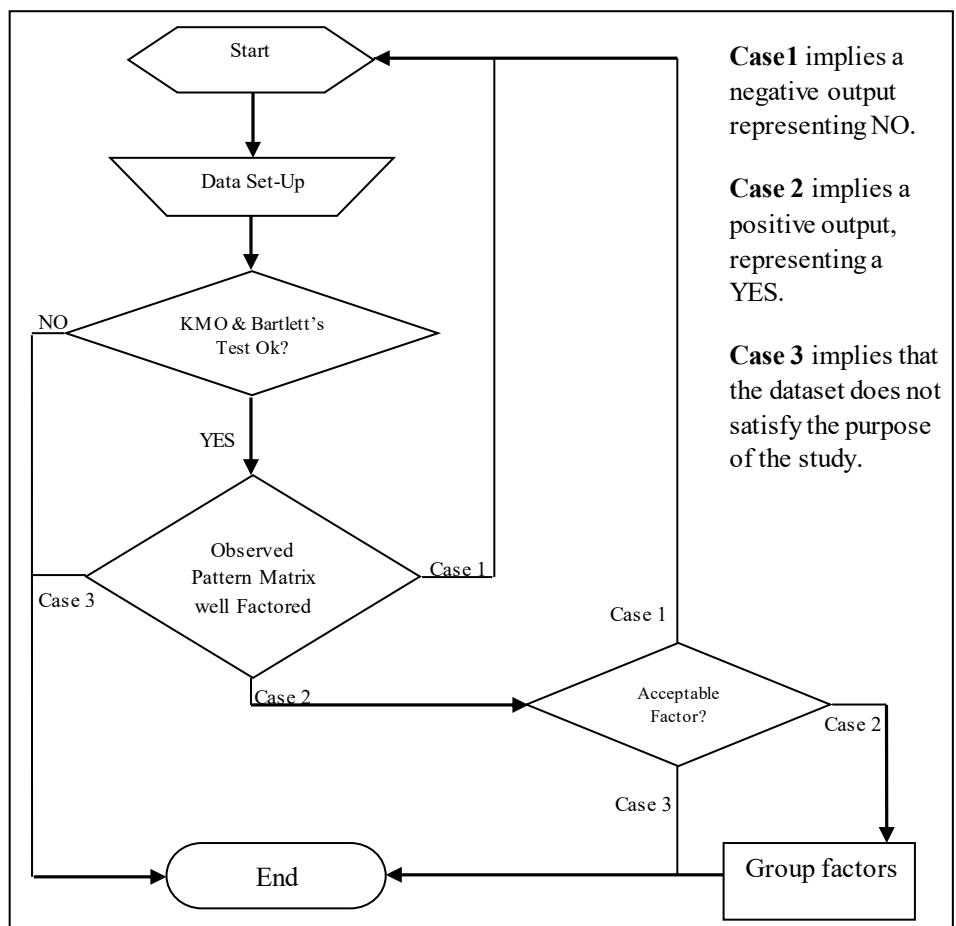

**Figure 5**: Flow Chart of Factor Analysis Process

The flow chart, as shown in Figure 5, starts with the preparation of the data for analysis as indicated by the symbol. A manual process of identification of observable variables proceeds the preparation phase, as indicated by the symbols. This is then followed a decision on output of KMO and Bartlett's test. Further decision on the output of the KMO and Bartlett's test is carried-out through observation and processing of the pattern matrix. Case 1 indicates that the desired output criteria are not met, case 2 indicate that the desired output are satisfied while case 3 indicates that the desired output cannot be satisfied by the dataset. Upon a desired factor output, the factors are then grouped to reflect the result. The output of the pattern matrix using a principal axis factoring and Promax with Kaiser Normalization is shown in Table 7.

**Table 7**: Pattern Matrix [a]

| Observable Variable | Factor | | | | |
|---|---|---|---|---|---|
| | 1 | 2 | 3 | 4 | 5 |
| Employed Staff 1 | .722 | | | | |
| Employed Staff 2 | .812 | | | | |
| Employed Staff 3 | .578 | | | | |
| Employed Staff 4 | .724 | | | | |
| contract Staff 1 | .542 | | | .277 | |
| Contract Staff 2 | .747 | | | | |
| Contract Staff 3 | .378 | .288 | | | |
| sacked Staff 1 | | | | .591 | |
| Sacked Staff 2 | .273 | | | .572 | |
| Sacked Staff 3 | | | | .694 | |
| Sacked Staff 4 | .303 | | | .550 | |
| Sacked Staff 5 | | | | .637 | |
| Affiliates 2 | | | .542 | | |
| Affiliates 3 | | | .605 | | |
| Affiliates 4 | | | .508 | | |
| Affiliates 5 | | | .884 | | |
| Affiliates 6 | | | .890 | | |
| Multiple access | | .646 | | | |
| Multiple access 2 | .217 | .574 | | -.258 | |
| Multiple Access 3 | .241 | .688 | | | |
| Multiple Access 4 | | .327 | | .246 | |
| Dual Knowledge | -.408 | .207 | | .396 | |
| Dual Knowledge 2 | | .703 | | | |
| Dual Knowledge4 | | .679 | | | |
| Multiple knowledge | | .773 | | | |
| Multiple Knowledge 2 | | .742 | | | |
| Triple Knowledge | | | | .207 | .633 |
| Triple Knowledge 2 | | | | | .945 |
| triple Knowledge 3 | | | | | .769 |
| Extraction Method: Principal Axis Factoring Rotation Method: Promax with Kaiser Normalization | | | | | |
| [a.] Rotation converged in 7 iterations. | | | | | |

Six factors were initially observed. However, a synopsis of the survey instrument using the principle of interpretability of factors, and theoretical expectation of number of constructs (Costello and Osborne 2005) as described in the outcome of the auto-covariance matrix presented in Table 2, this study adopted a five-factor classification system. Cross-loading problem was observed in the factors. Cross-loading occurs when a single observable variable is loaded by different factors. In order achieve better factorization, this study eliminated observable variables with variance cross-factor loading of more than 0.2 value.

However, after the deletion (approximate of 14% of the total observable), the study decided to terminate the elimination process. Moreover, at the point, the pattern matrix reflects the distinction among the factors, as shown in Table 7. For easy description, the factors are relabelled to reflect the underlying characteristics of the factors as detailed in Table 8. The relabelled factors are described as construct in the proceeding sections.

**Table 8**: Model Construct Description

| Factor | Underlying characteristics | Construct Description | Label |
|---|---|---|---|
| 1 | Describes perceived employed staff in an organization | Employed staff | ES |
| 2 | Employed staff with knowledge of multiple department in an organization | Dual Knowledge Staff | DK |
| 3 | Affiliates of an employed or ex-employed staff | Affiliate of employed staff | AFF |
| 4 | Describes subject perceived to be an ex-employee of the organization | Ex-Employed Staff | SS |
| 5 | Describes subjects with knowledge of more than two department in an organization | Triple Knowledge staff | TK |

## 6. Model Design And Validation

This study adopts the sequential procedure in SPSS-AMOS tool defined in Figure 6. The construct defined in Table 8 serves as the input for this phase of the study. The flow-chart for this phase is presented in Figure 7.

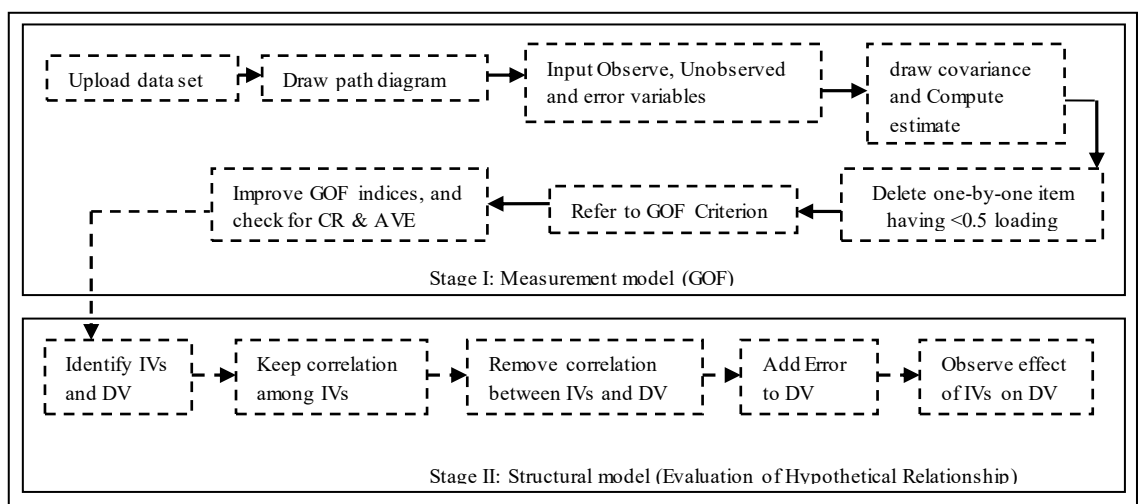

**Figure 6**: Model testing and hypothesizing process

As shown in Figure 6, the designed model comprises a measurement model, and a structural model. The measurement model examines the dataset for the fabricated constructs using multiple fit indices for goodness of fit (GOF). These indices include root mean square error of approximation (RMSEA), Comparative fit indices (CFI), ratio of Chi-square and degree of freedom (CMIN/DF), Composite reliability (CR), and average variance explained (AVE). The choice of these indices is based on the recommendation in (Hair, et al., 2010) and (Costello and Osborne 2005). Measurements are based on the thumb rule for each of the GOF indices, as shown in Table 8.

**Table 9**: Criterion for GOF (Source Hair et. al. 2010)

| Indices | N≥250, m≤30 |
| --- | --- |
| CMIN/DF | ≤3.00 |
| CFI | ≥0.92 |
| RMSEA | ≤0.07 |
| AVE | ≥0.5 |
| CR | ≥0.7 |

N= number of sample size
m= number of observable variable

Following the flow-chart in Figure 7, the model was observed to achieve a substantial goodness of fit (GOF) criterion. For the first iteration process as shown in Table 9, the covariance of all the constructs (at this phase, the constructs are generally classified without distinction on latent or independent variable).

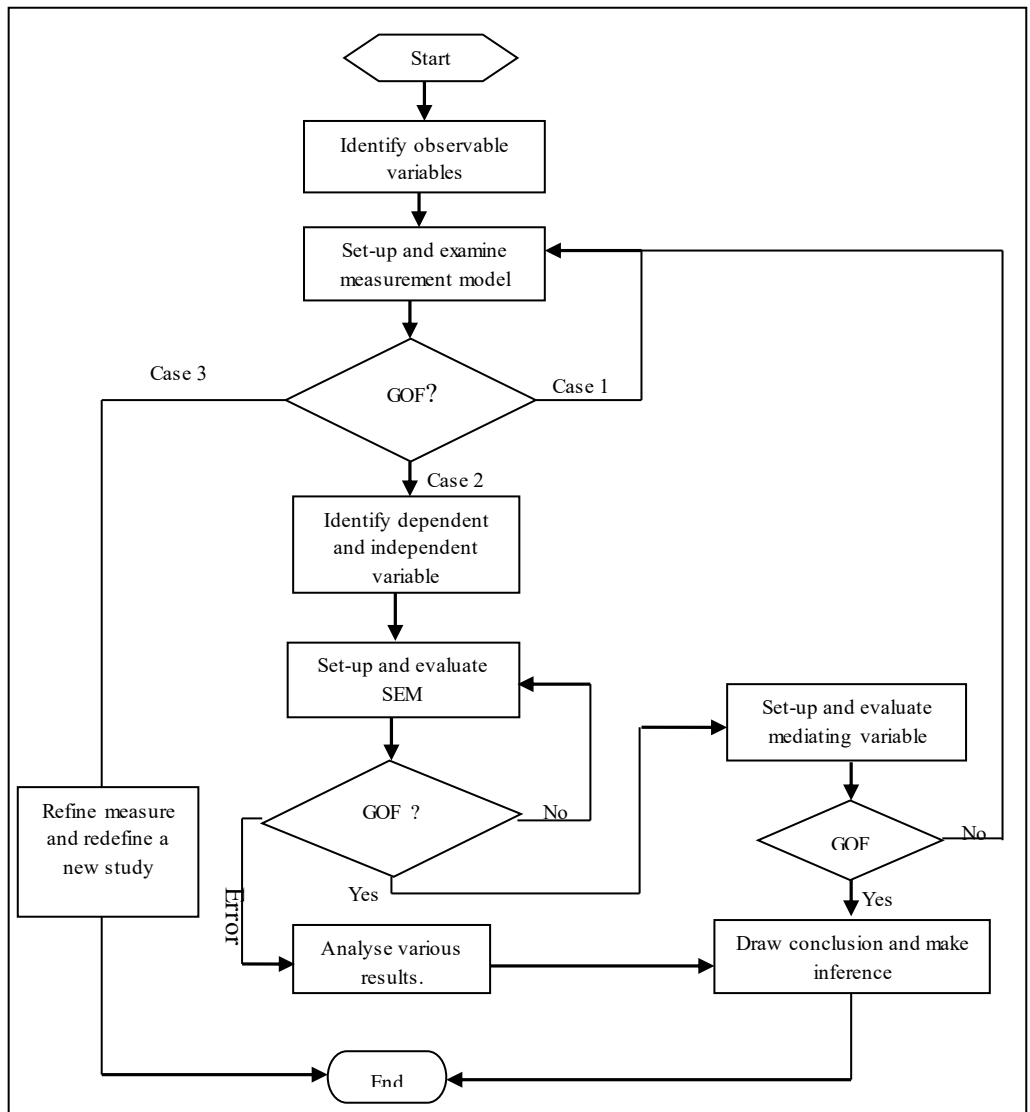

**Figure 7**: Flow-Chart for Proposed Model

From the observation, the CMIN/DF criteria of ≤3.00 was achieved as well as the composite reliability (CR). However, other criterions were not achieved. To improve the fitness of the model, unobservable variables possessing highest regression coefficient weight, which falls within the same constructs, are correlated. Thus, the second through the fourth iteration process of the correlates unobservable variables e8 & e9, e5 & e6, and e22 & e23 respectively. After these iterations, the convergence validity measure (Average variance explained: AVE) of the model was observed to fall within the threshold of value > 0.5.

**Table 10**: Goodness of Fit Indices for Measurement Model

| No | Stages | CMIN/DF | CFI | RMSEA | AVE | CR | Description/Outcome |
|---|---|---|---|---|---|---|---|
| 1 | Construct covariance | 2.697 | 0.890 | 0.072 | SS < 0.5 others>0.5 | All > 0.7 | Measurement model fit for some indices |
| 2 | e8↔e9 | 2.635 | 0.895 | 0.071 | SS < 0.5 | All>0.7 | AVE for SS = 0.482 |
| 3 | e5↔e6 | 2.540 | 0.901 | 0.069 | SS < 0.5 | All>0.7 | CR for all greater than 0.87 |
| 4 | e22↔e23 | 2.330 | 0.915 | 0.064 | SS < 0.5 | All>0.7 | |
| 5 | Delete SS2 | 2.378 | 0.915 | 0.065 | SS < 0.5 | All>0.7 | To improve AVE, remove factor with least regression weight in the model. AVE for SS = 0.496, CR for all greater than 0.754 |
| 6 | e1↔e2 | 2.303 | 0.923 | 0.063 | SS < 0.5 | All > 0.7 | AVE for SS = 0.496, |
| 7 | e18↔e16 | 2.259 | 0.923 | 0.062 | SS < 0.5 | All > 0.7 | CR for all greater than 0.754 |
| 8 | E14↔e15 | 2.150 | 0.930 | 0.060 | SS < 0.5 | All > 0.7 | |
| 9 | E2↔e6 | 2.077 | 0.934 | 0.058 | SS < 0.5 | All > 0.7 | |
| 10 | Delete SS6 | 2.065 | 0.939 | 0.057 | All > 0.5 | All> 0.77 | AVE for all greater than 0.5 CR for all greater than 0.754 No validity and reliability concerns. Model fit is good |

↔ Indicates correlation.

Moreover, observable variable SS2 was observed to have regression weight lesser than 0.5. Thus, SS2 was removed from the model. Further iterations were carried-out as detailed in Table 10, until an acceptable statistical validity (adequate model fit and construct validity) was achieved. The path diagram of the measurement model is presented in Figure 8. After the 10$^{th}$ iteration process, the measurement model was observed to attain statistical reliability and validity.

This implies therefore that these sets of constructs can be used to study relationship between employees of an organization and affiliates, as well as ex-employees, which constitutes the five constructs carved-out in this study. In order to carry-out the structural theory test, stage II of Figure 6 was implemented. It begins with the identification of independent and dependent variables. With reference to the equations in Appendix A, an employed staff through interaction with other employed staff within and without same department, can gain knowledge which could be adequate for access knowledge and or right. Similarly, interaction between employed staff and affiliates of an employed staff as well as ex-employed staff could yield substantial information for access knowledge and or right. Hence, this study identifies these constructs thus:

i. Employed staff (ES): Independent construct
ii. Dual Knowledge Staff (DK): dependent construct
iii. Ex-employed staff (SS): dependent construct
iv. Affiliate of employed/ex-employed staff (AFF): dependent construct
v. Triple Knowledge Staff (TK): dependent construct

DK, SS, TK, and AFF are dependent on ES. Therefore, ES is an exogenous construct (predictors), while DK, SS, TK, and AFF are endogenous construct (outcome). The hypothesis to validate does as follows:

i. H1 (ES→DK): the overall interaction between employee, either in same of different department could be substantial enough to the degree of the existence of dual knowledge staff, either through direct or indirect relationship
ii. H2 (ES→SS): the overall interaction between ES and SS could be positive such that there exist a common knowledge among them through direct or indirect relationship
iii. H3 (ES→AFF): overall interaction between ES and AFF either directly or directly could be positive such that there exist the possibilities of AFF gaining access knowledge and or right
iv. H4 (ES→TK): overall interaction among ESs in same or different department as well as among AFFs, could be substantial such that there exist the possibilities of a TK

IV/DV relationship was introduced as shown in Figure 8, and correlations between the DVs were deleted. In order to evaluate the model and test the hypothesis, we ensured that the direct and indirect relations among construct of interest was added, by connecting single arrow from the predictor to the various outcomes. Furthermore, we also observed the relationship such that the two other endogenous constructs (SS, DK) serves as predictor to the other endogenous constructs (AFF, TK). This was observed to yield a significant regression coefficient in the model. As shown in Figure 8, both direct and indirect effects are observed on the construct. H1 hypothesize the direct and indirect relationship between ES and DK, H2 hypothesize the direct relationship between ES and SS, H3 hypothesize the direct and indirect relationship between ES and AFF, while H4 hypothesize the direct and indirect relationship between ES and TK.

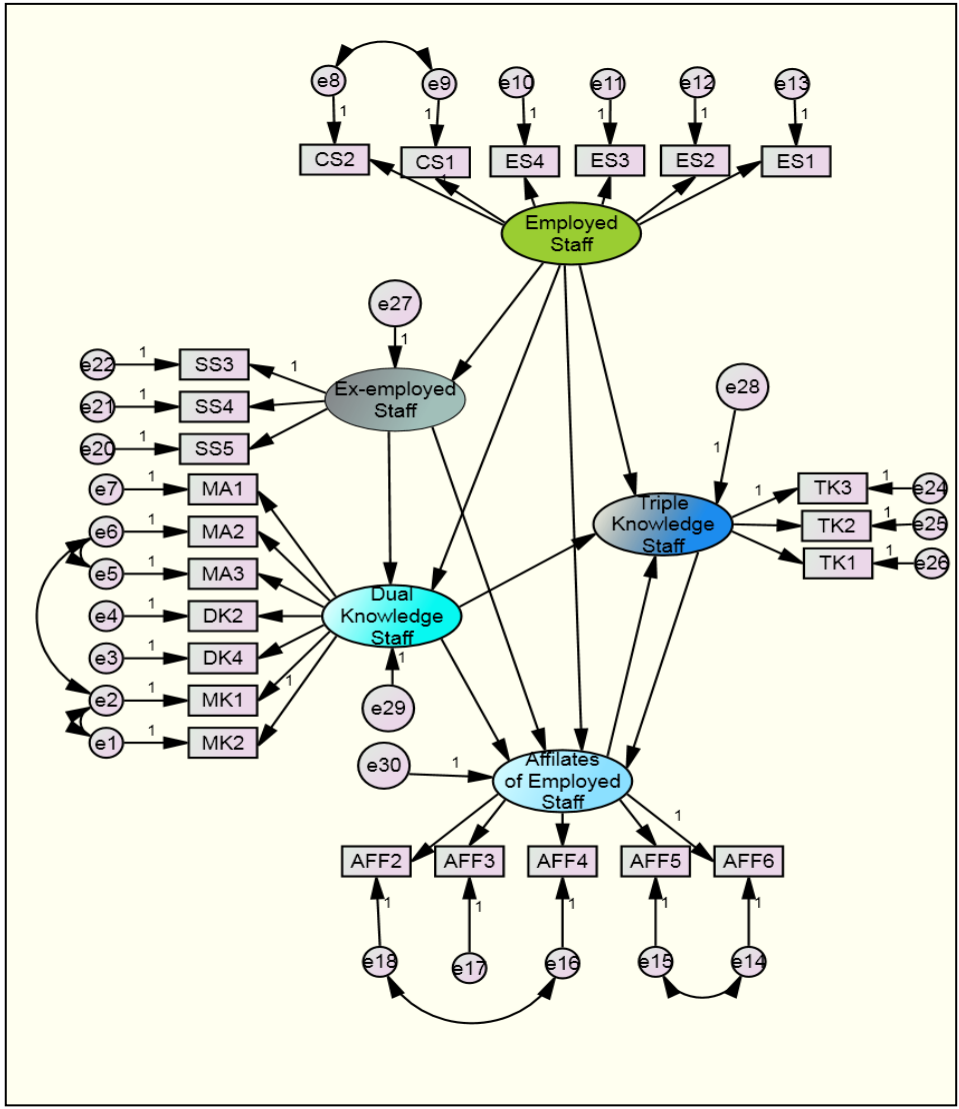

**Figure 8**: Path Diagram of Structural Equation Modelling for Insider Taxonomy

ES has three structural path of indirect relationship to AFF, and two structural path of indirect relationship with TK as expected from the classification matrix in Table 2, thus establishing triple knowledge possibilities. The indirect relationship between ES and DK further supports the theory of the outcome matrix in Table 1. Table 11 shows the overall correlation estimate of the relationship between the constructs in the model. Table 12 shows the comparison between the structural model and the measurement model.

**Table 11**: Standardized Total Effects

|  | Employed_Staff | Ex-employed_Staff | Dual_Knowledge_Staff | Triple_Knowledge_Staff | Affilates_of_Employed_Staff |
|---|---|---|---|---|---|
| Ex-employed_Staff | .528 | .000 | .000 | .000 | .000 |
| Dual_Knowledge_Staff | .704 | .298 | .000 | .000 | .000 |
| Triple_Knowledge_Staff | .249 | .432 | .543 | -.076 | .431 |
| Affilates_of_Employed_Staff | .262 | .715 | .460 | -.162 | -.076 |

There is significant statistical correlation among all the constructs as shown in Table 11, indicating the acceptance of the alternate hypotheses and rejection of the null hypotheses. However, statistical correlation peaks at ES →DK, and troughs at ES→TK, describing the possible level of interaction between the constructs.

**Table 12**: Comparison Indices between Structural Model for Insider taxonomy and the CFA Measurement Model

| GOF INDICES | Structural model | CFA model |
|---|---|---|
| CMIN | 487.434 | 487.434 |
| DF | 236 | 236 |
| CMIN/DF | 2.065 | 2.065 |
| P | 0.000 | 0.000 |
| RMSEA | 0.057 | 0.057 |
| CFI | 0.939 | 0.939 |

As shown in Table 12, there is no statistical variance between the two models. This shows that the model provides a good overall model fit which is constituent for both models. It also implies there are no interpretational confounding errors. Interpretational confounding reveals structural misspecification, as well as measurement error. Since there is no variance between the models, it can be said that the model perfectly fit for insider taxonomy.

## 6.1 Result Discussion and Limitation

Our result shows that there is no clear-cut distinction between an employer and a contract staff in an organization. This result therefore supports the recommendation in (Neumann, 2010; Sarkar, 2010) which identifies contract staff as a potential employee, and should be address as such.

From the result shown in Table 11, various observations can be inferred about the relationship between the dependent variable and the independent variables. First off, there is a statistically significant relationship between the independent construct (ES), and the dependent constructs (DK, SS, AFF, TK). Hypothesis H1, H2, H3, H4 therefore holds true for each of the relationships, thus establishing the basis for which DK, SS, AFF, TK can be classified in line with ES as the composition of insider to an organization. This study therefore rejects the null hypothesis of the models, favouring the acceptance of the alternate hypothesis, which describes the relationship between the constructs. Moreover, the coefficient of correlation between the independent variable and the dependent variable (0.528, 0.704, 0.249 and 0.262) depicts the operational reality of humans/subjects that contribute to the day-to-day activities of an organization. This explains the need for the evolved paradigm of insider taxonomy, intention dissection notwithstanding. From the result therefore, it follows that an insider classification system especially where insider misuse is involved, requires a thorough consideration of the possible connection between different subjects, as well as between subjects and their possible affiliates. It may be possible that an affiliate perpetuate a particular act with the access right of a subject, with or without the knowledge of the subject in question. Such clarification would require a proper dissection of all the affiliation related to each subjects. Furthermore, the interoperability between subjects of different classes and clearance level may generate useful artefacts, which could have been otherwise overlooked. Appropriating this result into organizational security framework could be a possible way of identifying possible breaches, and curbing insider misuse possibilities. This research thus fills the gap of identifying the operational composition of organization's day-to-day activities. However, this research is limited in its incapacity to delineate the perspective of insider from each categories of organization. Furthermore, it failed to provide insight into insider perception from societal differences perspective. This is anchored on the premise that (if) human interaction forms the cardinal for which holistic taxonomy of insider can be viewed, then it surmise to state that interaction differs from one society to another.

## 7. Conclusion and Future Work

This study presents the result of a conceptual model for insider taxonomy from the evolving paradigm of classical insider description. Using questionnaire instrument, from three categories of organization, this study models an outcome matrix of insider taxonomy. Statistical analysis tools, and structural equation modelling tool was adopted for analysis and modelling process. The sample provided the minimal requirement for which generalized findings can be extracted. The result reveals the "real operational" description of insider constituents, as against the subject-object description. From the result, is it observed that there is a statistical significance between the designed variables which defines who an insider is. This result can be applied for investigation process of insider crime and security related alerts. Furthermore, this result can be applied in staff training as well as implemented in organizational policies to manage the effectiveness of staffs, evaluate staff propensity to malicious intention, or provide interactive policies for effectiveness. This can be done by reviewing the various dependencies and level of interaction between each identified subject.

This study is part of an on-going research on insider taxonomy in relation to misuse investigation. As part of the continuing work, this study intends to further examine the variability in description of insider from the three distinct organizations in order to understand the level of interaction between the identified variables. Further studies on insider taxonomy based on societal differences will greatly improve this paradigm of insider definition.

**Appendix A**

Social interaction, relationship and interdependence theory suits this study. This theory "presents a logical analysis of the structure of interpersonal relationship", thus offers a conceptual framework for interpersonal situational analysis, as shown in equations shown below

$$B = f(P, E)$$

where B signifies behavior, $f(P, E)$ represents function of the property of the person ($P$), and the environment ($E$) in context.

$$I = f(S, A, B)$$

Interaction ($I$) is a function of social situation ($S$) between persons (A) and (B). "The option and outcome of interaction can be represented using a tool from the classic game theory: the outcome matrix". An outcome matrix describes interdependence pattern among people, thus useful in describing social situation, in that it describes the intricacies and degree of interaction. This follows suit with Locard's exchange principle of exchange theory of transference.